\documentclass[a4paper]{article}

\usepackage{INTERSPEECH2021}
\usepackage{footnote}
\usepackage{cite}
\usepackage{pifont}
\title{An Integrated Framework for Two-pass Personalized Voice Trigger}
\name{Dexin Liao$^1$, Jing Li$^1$, Yiming Zhi$^1$, Song Li$^2$, Qingyang Hong$^1$, Lin Li$^2$}
\address{
  $^1$School of Informatics, Xiamen University, China\\
  $^2$School of Electronic Science and Engineering, Xiamen University, China}
\email{qyhong@xmu.edu.cn, lilin@xmu.edu.cn}

\begin{document}

\maketitle
% to minimise a phonetic connectionist temporal classification (CTC) loss/score calculated from the miss rate and the false alarm rate can reach about 0.14 on evaluation set in task 1
\begin{abstract}
In this paper, we present the XMUSPEECH system for Task 1 of 2020 Personalized Voice Trigger Challenge (PVTC2020). Task 1 is a joint wake-up word detection with speaker verification on close talking data. The whole system consists of a keyword spotting (KWS) sub-system and a speaker verification (SV) sub-system. For the KWS system, we applied a Temporal Depthwise Separable Convolution Residual Network (TDSC-ResNet) to improve the system's performance. For the SV system, we proposed a multi-task learning network, where phonetic branch is trained with the character label of the utterance, and speaker branch is trained with the label of the speaker. Phonetic branch is optimized with connectionist temporal classification (CTC) loss, which is treated as an auxiliary module for speaker branch. Experiments show that our system gets significant improvements compared with baseline system.
\end{abstract}
\noindent\textbf{Index Terms}: Keyword spotting, Speaker verification, TDSC-ResNet, Multi-task, CTC
%The 2020 Personalized Voice Trigger Challenge (PVTC2020) includes two tasks, joint wake-up word detection with speaker verification on close talking data (task 1) and joint wake-up detection with speaker verification on far-filed multi-channel microphone array data (task 2).
%Although \cite{sigtia2020multi,jung2020multi} demonstrated that it is possible for a single model to jointly optimize wake-up word detection and speaker verification
%The database XIAO-LEis recorded under the smart home scenario.
% Typically SV task needs lager networks than KWS task, so verifying identity of the audio which doesn't contain keyword will lead to unnecessary computation and high latency. However,

\section{Introduction}
Voice wake-up is becoming popular in people everyday life as smart devices are widely used. Usually, detecting a trigger phrase including two stages, the first is a KWS system to detect the prefixed keywords, and the second is a SV system to conform the identity of the speaker. PVTC2020 \cite{jia20212020} includes joint wake-up word detection with speaker verification on close talking data (task 1) and joint wake-up detection with speaker verification on far-field multi-channel microphone array data (task 2). More details about the dataset, task and evaluation metrics refer to \cite{jia20212020}. In this paper, we mainly focus on task 1. From a practical standpoint, the whole system should satisfy the requirements of low latency. Typically, SV task needs lager networks than KWS task, an always-on SV model will lead to unnecessary computation and high latency. So we employed a cascaded 2-stage architecture, where a low cost KWS system decides whether triggered and then the keyword segment is fed to a larger SV system to do further identification.
%determine whether it is uttered by the enrolled target speaker.

Currently, the Deep-KWS \cite{chen2014small} system has become more popular for its simpler implementation and higher accuracy. Many Convolutional Neural Networks (CNNs) architectures \cite{tang2018deep,coucke2019efficient,xu2020depthwise,choi2019temporal,yang2020multi,li2020small,higuchi2020stacked} have been explored for KWS system to achieve high accuracy with limited model size and computational resources. Temporal Convolution (TConv) \cite{choi2019temporal} extracts the whole level frequency features to address the problem that conventional 2D CNNs struggle with concentrating the relation between high and low frequencies. Depthwise Separable Convolution Neural Network \cite{howard2017mobilenets} considers the channel realm and space realm separately, reduces the number of the parameters of the standard convolution without significant performance degradation, and is well applied on KWS\cite{xu2020depthwise}. To achieve a fast and accurate KWS system, we applied TConv and depthwise separable convlolution on our Deep-KWS system.

In SV field, x-vector framework proposed by Snyder et al. \cite{snyder2017deep} is the most popular framework recently. X-vector can make full use of deep neural network and significantly improve the performance of SV systems. In recent years, multi-task (MT) frameworks have been applied to combine speech with speaker information to improve the performance of SV systems \cite{liu2018speaker,wang2019usage,zhao2019phone}. These works showed that we can add frame-level phonetic information learned from multitasking before the pooling layer. It can help the SV system to distinguish speaker-specific information more easily. \cite{sigtia2020multi,jung2020multi} attempted to jointly optimize KWS and SV with a single network. In the speaker dependent voice trigger task, the SV system aims to verify the speaker identity of utterance containing prefixed keywords. Due to the above, we reproduced the frame-level MT framework \cite{wang2019usage} to better utilize the phonetic information in SV system. However, this method needs to get the phoneme labels of the inputs on frame-level. Usually the alignment is generated by a well trained automatic speaker recognition (ASR) model. The procedure is complex and the accuracy of labels is subject to the performance of ASR model. What's more, the ASR model might have decoding failure when processing the audio which is domain mismatched or noisy. CTC \cite{graves2006connectionist} can automatically label unsegmented sequence data and has achieved great success in many end-to-end applications \cite{kim2017joint,karita2019improving,sainath2019two}. Motivated by \cite{sigtia2020multi}, we considered that CTC could be a simpler approach to embed frame-level phonetic information into SV system via MT learning. Thus, we proposed a multi-CTC x-vector network for PVTC2020 task 1 and significantly improved the performance compared with the official baseline.   
%has achieved great successful in many end-to-end applications \cite{kim2017joint,watanabe2017hybrid,rao2017exploring,karita2019improving,sainath2019two}.

%rao2017exploring,karita2019improving,watanabe2017hybrid

We demonstrate the whole system in Section 2. Experimental details are presented in Section 3, and results are reported and analyzed in Section 4. Finally, we conclude this work in Section 5. 

\section{System description}

\begin{figure}[t]
	\centering
	\includegraphics[width=2.5in]{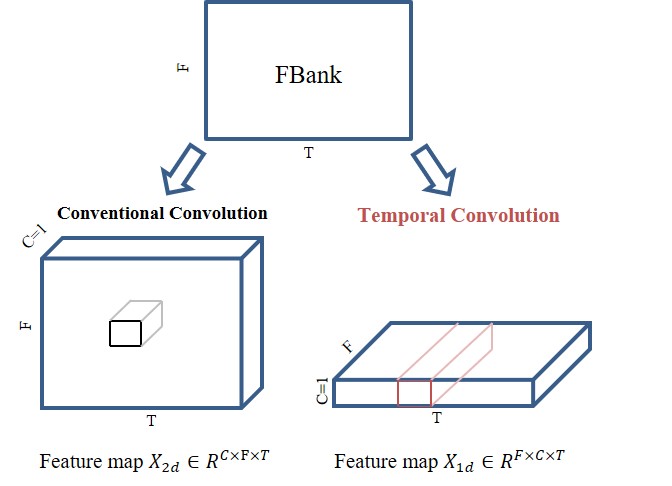}
	\caption{Schematic diagram of the difference between Conventional Convolution and Temporal Convolution.}
	\label{fig:Tconv}
\end{figure}
\subsection{KWS sub-system}
The KWS system consists of three modules:(i) a feature extraction module, (ii) a deep neural network, and (iii) a posterior handling module. The features are 80-dimensional log-mel filterbank (FBank) computed with a 25ms window and shifted every 10ms. And we apply a window of 40 frames, which contain context information of sub-keywords, to generate training samples as the input of the model. We choose CNN as the acoustic model of KWS system. In the posterior handling module, the sequence of acoustic features is projected to a posterior probabilities sequences of the sub-words after the acoustic model. We smooth the posteriors over a fixed window of size 50 by taking average, and the confidence score is computed within a sliding window of size 150. The method of confidence computation is proposed in \cite{liu2019loss}, in which the confidence score subject to the constraint that the sub-keywords are uttered in the same order as in the specified keyword. The system triggers whenever the confidence score is higher than a predefined threshold.
%we use the official pipeline which adopt the method proposed in \cite{liu2019loss}, in which the confidence score subject to the constraint that the sub-keywords are uttered in the same order as in the specified keyword.
\subsubsection{Temporal convolution}
 
The Temporal Convolution ResNet (TC-ResNet) model \cite{choi2019temporal} is composed of sequence of residual blocks which use one dimensional convolution named Temporal Convolution (TConv). We treat the input FBank as time series. Conventionally, the FBank features are denoted as $\mathbf{I} \in \mathbb{R}^{ F \times T}$, where F represents the dimension of FBank feature, and T denotes the number of frames. As shown in Figure~\ref{fig:Tconv}. Conventional CNNs reshape the input from $\mathbf{I}$ to  $\mathbf{X_{2d}} \in \mathbb{R}^{ C \times F \times T}$, where $C=1$ means the channel dimension. In TConv, the feature dimension is equal to the channels of the input feature map, so the input feature map $X$ is transformed to  $\mathbf{X_{1d}} \in \mathbb{R}^{ F \times C \times T }$, where $F=80$ and $C=1$. Therefore, all convolutions in the model are along the temporal dimension, avoiding stacking a large number of layers to capture high-level frequency features. In addition, TConv requires a smaller number of computations compared with the conventional 2d convolution. What's more, it can decrease the output feature map (i.e., the input feature map of the next layer) of the TConv and dramatically reduce the computations in the following layers.

\subsubsection{Model architecture overview}
\begin{figure}[t]
	\centering
	\includegraphics[width=\linewidth]{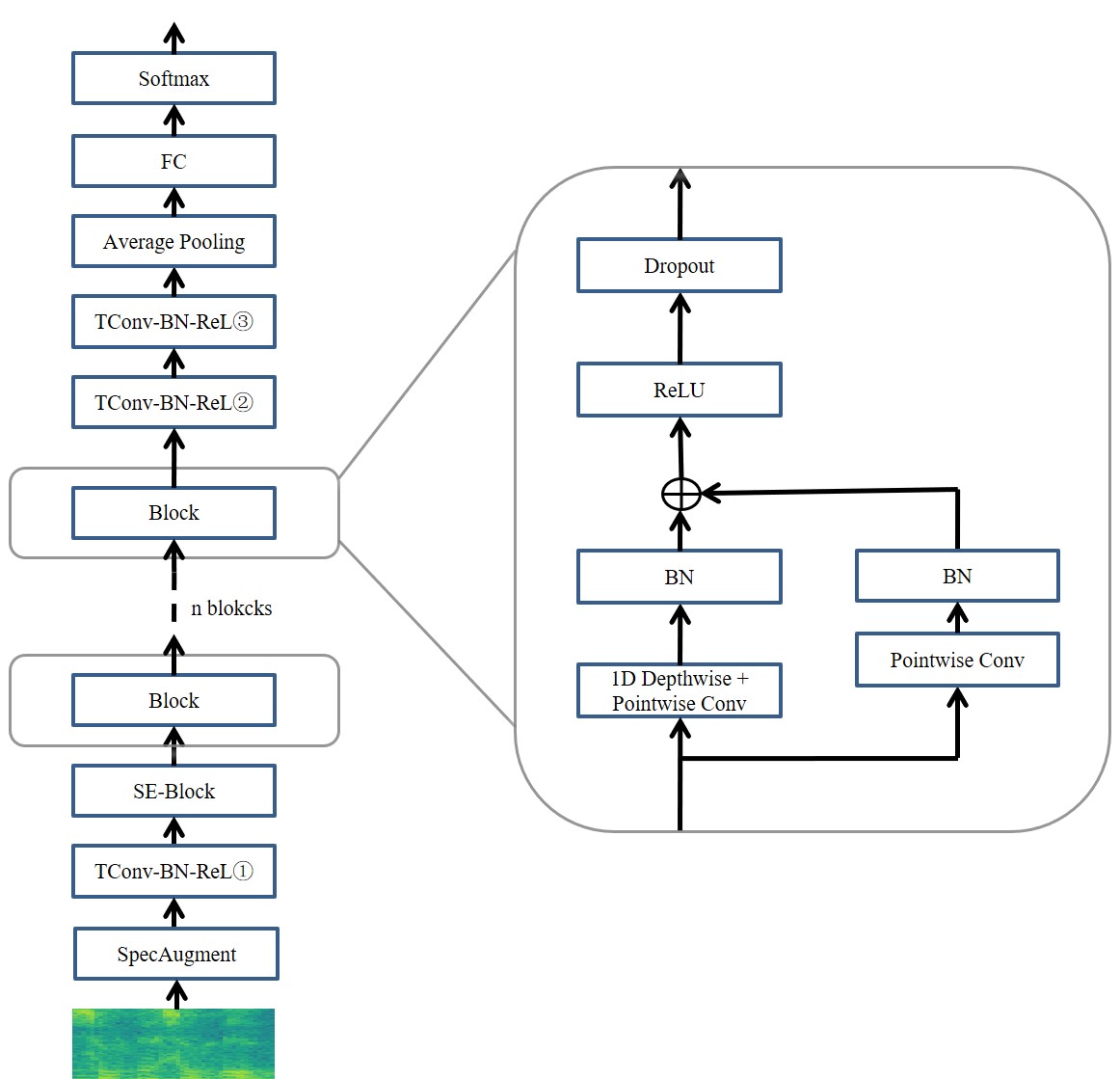}
	\caption{Schematic diagram of TDSC-ResNet.}
	\label{fig:speech_production}
\end{figure}
Our Temporal Depthwise Separable Convolution ResNet (TDSC-ResNet) is based on MatchboxNet \cite{majumdar2020matchboxnet}. As shown in Figure~\ref{fig:speech_production}, the model uses TConv and depthwise and pointwise convolution. We first apply SpecAugment \cite{park2019specaugment} on the bottom of the architecture thus enable the acoustic model to become more robust. After SpecAugment, the input feature map  $\mathbf{I} \in \mathbb{R}^{ F \times T}$ is transformed to $\mathbf{X_{1d}} \in \mathbb{R}^{ F \times 1 \times T }$ as illustrated in Sec. 2.1.1. Then TConv and Squeeze-and-Excitation (SE) \cite{hu2018squeeze} block can better accommodate feature maps. Inside the repeated blocks are separable convolution, 1$ \times $1 pointwise convolutions, batch norm, ReLU, and dropout (see the right part of Figure~\ref{fig:speech_production}), and we set the $n=3$ which means there is 3 blocks in the architecture. Finally we position two layers of additional TConv followed by an average pooling layer and then a fully connected layer and a softmax activation layer are applied to obtain the sub-words occurrence probability of the keyword. The detail settings of the model is listed in Table~\ref{tab:network_config}. The receptive field of the model is designed according to the length of input samples. 

\begin{table}[th]
	\caption{The configuration of TDSC-ResNet. The number after Block denotes the index of the repeated blocks.}
	\label{tab:network_config}
	\centering
	\footnotesize
	\resizebox{\linewidth}{!}{
		\begin{tabular}{lcc}

    		\toprule
			\textbf{Layers}               & \textbf{Kernel size}  & \textbf{Channels} \\ 
			\midrule
			TConv\ding{172}  & 1x6 & 128 \\
			Block\ding{172} Depthwise Conv  & 1x7 & 128 \\
			Block\ding{172} Pointwise Conv  & 1x1 & 64 \\
			Block\ding{173} Depthwise Conv  & 1x8 & 64 \\
			Block\ding{173} Pointwise Conv  & 1x1 & 64 \\
			Block\ding{174} Depthwise Conv  & 1x9 & 64 \\
			Block\ding{174} Pointwise Conv  & 1x1 & 64 \\
			TConv\ding{173}    & 1x14 & 128     \\
			TConv\ding{174} & 1x1 & 128     \\ 
			\bottomrule
		\end{tabular}
	}	
\end{table}

%\subsubsection{Posterior Handling}  
%The sequence of acoustic features is projected to a posterior probabilities sequences of the sub-words after the acoustic model. We smooth the posteriors over a fixed time window of size 40 by taking average, and the confidence score is computed within a sliding window of size 150. The method of confidence computation is proposed in \cite{liu2019loss}, in which the confidence score subject to the constraint that the sub-keywords are uttered in the same order as in the specified keyword. The system triggers whenever the confidence score is higher than a predefined threshold.
\subsection{SV sub-system}
The standard x-vector system supports the input to have fixed length. We first modified the x-vector system to allow various lengths input to better fit the CTC loss training. The feature is 40-dim MFCCs with 3-dim Pitch, 25ms frame length and 10ms shift. Cepstral mean-normalization
(CMN) with a sliding window of 3 seconds and voice active detection (VAD) are also applied. In the training phase, the training data contains keyword and other contents. But in the evaluation phase, the SV system extract discriminative speaker embeddings of the keyword segments, which are much shorter than training data. Hence, we optimize the model through two steps. First the model is trained with all the training data. Then only the keyword segments are used to fine-tune the model again. SpecAugment is applied during training. In enrollment stage, the average vector of the three utterances' embedding extracted from the target speaker is saved as the enrollment speaker embedding vector. 
\subsubsection{Connectionist temporal classifier}
Unlike cross entropy, in which the alignments between input and target sequences should be given, the CTC loss automatically learn the alignments. Suppose there are $ N $ modeling label units, the softmax layer before CTC consists of $ N $ units and a blank unit. The introduction of blank unit enable outputs to define the probabilities of all possible alignments of the input sequence. Given a T frames utterance, the CTC path $ \pi=(\pi_0,...,\pi_{T-1}) $ contains repetitions of no-blank labels and the blank unit. After a many-to-one mapping function $\beta$, which means removing the repeated labels and blanks, the CTC path can be mapped to the corresponding label sequence (e.g. $\beta$(a-ab-) = $\beta$(-aa- -abb) = aab). Summing all the possibilities of CTC paths has heavy computing burden, and a forward-backward dynamic programming algorithm \cite{graves2006connectionist} was proposed to solve this problem. Given the label sequence, CTC naturally tends to align each label prediction. In this way, the sequence of frame-level posterior probability distribution before CTC loss is similar to that before CE loss, while the later requires a predetermined alignment. 
 
\subsubsection{The frame-level multi-task learning}
\begin{figure}[t]
	\centering
	\includegraphics[scale=0.35]{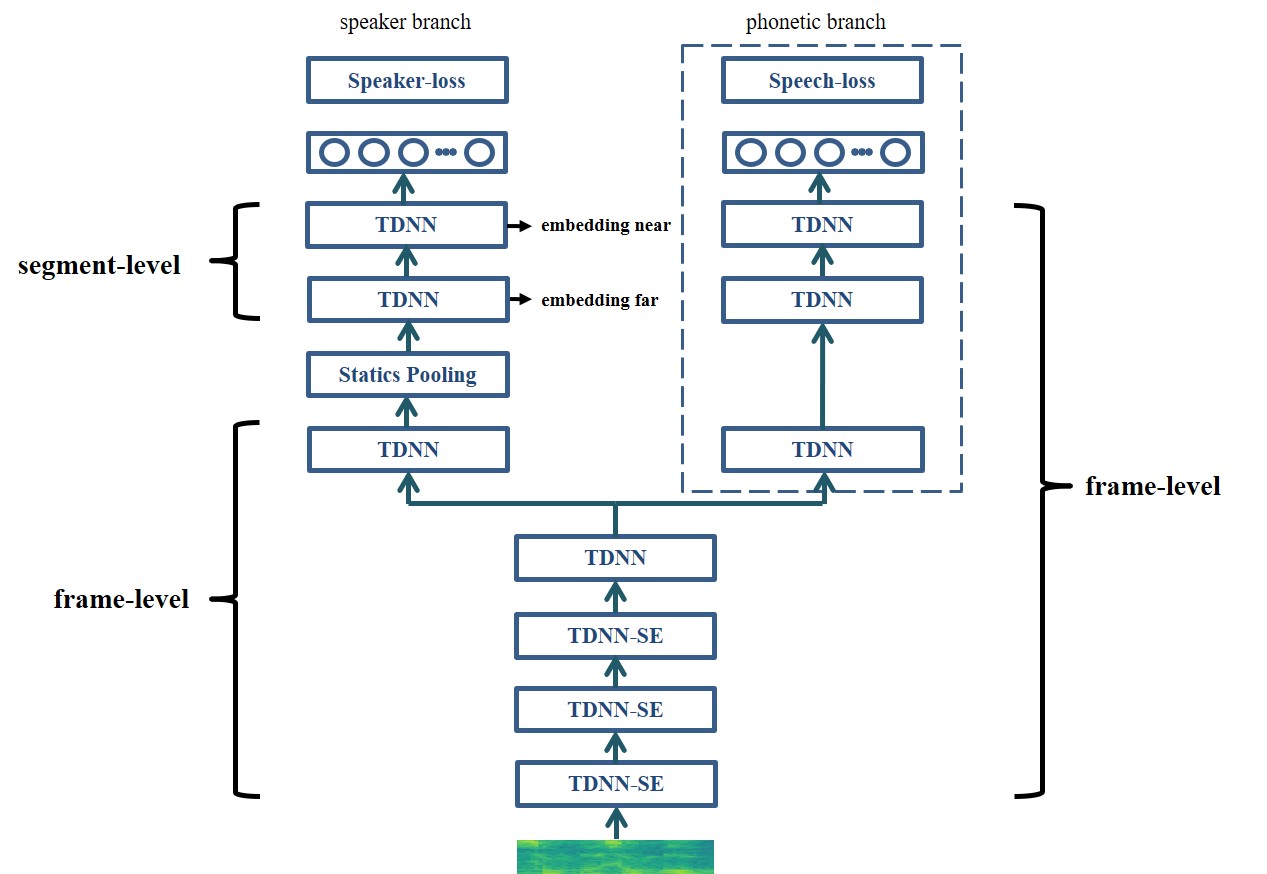}
	\caption{Frame-level phonetic MT x-vector framework.}
	\label{fig:multi-task}
\end{figure}
We adopt time delay neural network (TDNN) with SE-block as the speaker embedding extractor, which is optimized by AAM-Softmax \cite{deng2019arcface} loss. The detailed configuration is shown in Table~\ref{tab:x-vector}. The MT system is depicted in Figure~\ref{fig:multi-task} and the configuration of TDNNs in phonetic branch is the same as that in speaker part. Unlike the MT system in \cite{sigtia2020multi}, the phonetic branch and the speaker branch in our MT system are trained jointly by a total loss function (\ref{eq1}) for each examples, and the phonetic branch is more inclined to assist SV branch to achieve a better performance. After training, the phonetic part is removed from the model. For the back-end, the $ embedding \ far $ in Figure~\ref{fig:multi-task} is for PLDA scoring and $ embedding \ near $ for cosine scoring. PLDA model is trained corresponding to the train set. Submean and normalization are applied during scoring.
\begin{equation}
	L_{loss} =  L_{speaker} + \alpha L_{speech}
	\label{eq1}
\end{equation} 

\begin{table}[]
	\caption{The configuration of the speaker embedding extractor.}
	\label{tab:x-vector}	
	\centering
	\footnotesize
	\resizebox{\linewidth}{!}{
		\begin{tabular}{ccccc}
			\toprule
			\textbf{Layer-index}	&\textbf{Layer}     & \textbf{Context}          &\textbf{SE-ratio}         & \textbf{Size}           \\ 
			\midrule
			1			&TDNN-SE    & \{t-2:t+2\}      &4         & 512            \\
			2			&TDNN-SE    & \{t - 2,t,t + 2\}&8         & 512            \\
			3			&TDNN-SE    & \{t - 3,t,t + 3\}&8         & 512            \\
			4			&TDNN    & \{t\}            &         & 512            \\
			5			&TDNN    & \{t\}            &         & 1500           \\
			6			&Stats pooling & [0,T]          &         & 2 x 1500  \\
			7			&TDNN  & \{0\}            &         & 512            \\
			8			&TDNN  & \{0\}            &         & 512            \\
			9			&AAM-Softmax   & \{0\}        &         & Num.spk \\
			\bottomrule
		\end{tabular}
		
	}
\end{table}

For the multi-Cross-Entropy x-vector (MT-CE-XV) system, the phonetic branch in Figure~\ref{fig:multi-task} is trained by CE loss. To get the aligned label, we used the PVTC training data to train a lattice-free MMI based ASR \cite{povey2016purely} refer to Kaldi \cite{povey2011the} recipe\footnote{https://github.com/kaldi-asr/kaldi/blob/master/egs/aishell/s5/run.sh}. Then, the frame-level phoneme alignment is obtained through the ASR model. Assume an utterance with T frames, the $ L_{speech} $ in (\ref{eq1}) is defined as (\ref{eq2})
\begin{equation}
	L_{speech} = \frac{1}{T} \sum_{t=1}^{T}CE(O^p_t,Y^p_t)
	\label{eq2}
\end{equation} 
where $ O^p_t $ denotes the output of the phonetic branch at time \textit{t}, $ Y^p_t $ is its corresponding frame-level phoneme label. 

For the multi-CTC x-vector (MT-CTC-XV) system, the $ L_{speech} $ is (\ref{eq3})
\begin{equation}
	L_{speech} =  L^\pi_{CTC}(Y_l)
	\label{eq3}
\end{equation} 
where $ \pi $ is the possibility distribution permutation before the CTC and $ Y_l $ is the label sequence.

A multi-CE-CTC x-vector (MT-CTC-CE-XV) system is a combination of CE and CTC loss in MT learning. That is, there are two phonetic branches, one for CE loss and another for CTC. Here the $ L_{speech} $ is (\ref{eq4})
\begin{equation}
	L_{speech} = \frac{1}{T} \sum_{t=1}^{T}CE(O^p_t,Y^p_t) + L^\pi_{CTC}(Y_l)
	\label{eq4}
\end{equation} 

\begin{table*}[ht]
	\caption{Results of different systems.}
	\footnotesize
	\label{tab:results}
	\centering
	\resizebox{\linewidth}{!}{
		\begin{tabular}{lllccll}
			\toprule
			\textbf{System index}  & \textbf{KWS system}  & \textbf{SV system}  & \textbf{SV-dev-EER\% }
			& \textbf{SV-dev-minDCF} & $ \bm{score_{dev}} $ & $ \bm{score_{test}} $    \\ 
			\midrule
			S1 & KWS-baseline\cite{jia20212020} & SV-baseline\cite{jia20212020} & 1.32 
			& 0.16  & 0.10  & 0.37 \\
			S2 \cite{leaderboard}  &            &   &  
			&   &   & 0.075 \\
			S3 \cite{hou2021npu} &  &  &  
			&  & 0.042  & 0.081 \\
			\midrule			
			S4 & XMU-KWS & SV-baseline & 1.368 & 0.162 & 0.090 & 0.161 \\
			S5 & XMU-KWS & x-vector & 1.572 & 0.195 & 0.099 & 0.189 \\
			S6 & XMU-KWS & MT-CE-XV & 1.159 & 0.152 & 0.073 & 0.168 \\
			S7 & XMU-KWS & MT-CTC-XV-phoneme   & 1.291 & 0.143 & 0.078 & 0.162 \\
			S8 & XMU-KWS & MT-CTC-XV-char   & \textbf{0.974} & \textbf{0.112} & \textbf{0.062} & \textbf{0.161} \\
			S9 & XMU-KWS & MT-CTC-CE-XV   & 1.237 & 0.123 & 0.073 & 0.168 \\
			S10 & XMU-KWS & MT-CTC-XV-char + score-fusion with PLDA   & \textbf{0.808} & \textbf{0.086} & \textbf{0.051} & \textbf{0.140} \\
			S11  &         & + Data augmentation                      & 0.786          &  0.083         &
			0.049     & 0.126 \\
			S12    &       & + model-fusion with SV-baseline      & \textbf{0.570} &\textbf{0.061}         &  \textbf{0.038}     &\textbf{0.098}  \\
			\bottomrule
		\end{tabular}	
	}
\end{table*}
\section{Experimental setup}

%toolkits:ASV-Subtool
All of the models are implemented based on our open source toolkits:ASV-Subtool \cite{tong2021asv}. In the KWS system, The validation set is randomly selected 10 out of 300 speakers' data in the training set. The keyword `xiao le xiao le' is split to three sub-keywords (i.e., `1\_xiao le', `2\_le xiao', `3\_xiao le'), and each sub-keyword contains an interval between Chinese characters inside the keyword. The labels for each input is determined by force-alignment with a large ASR system as the official pipeline. In the evaluation period, a simple VAD is applied to detect voice activity, and only the audio clips which trigger the KWS system are saved to be further transformed to speaker embeddings.

To train the SV system, the model is first pre-trained by the all training data and then fine-tuned by the keyword segments. Here are some details about training strategy:
% and the weight decay is set to 0.0003
\begin{itemize}
\item KWS: Stochastic gradient descent (SGD) is used with a batch size of 128 for 100 epochs. The initial learning rate is set as 0.01 and decreases by a factor of 0.5 when the model reaches a valid loss plateau. Early stopping is employed when the valid loss is not decreasing. 	
\item Pre-train SV: The Radam  optimizer \cite{liu2020on} is used to optimize the models with a batch size of 32 for 20 epochs.
\item Fine-tune SV: SGD optimizer with a batch size of 128 for 20 epochs. The learning rate is initialized as 0.03 and decreases by a factor of 0.5 when the model reaches a valid loss plateau. 
\end{itemize}

\section{Results and analysis}

\begin{table}[ht]
	\caption{Performance of the KWS models on dev set (the false rejection (FR) rate [\%] under one false alarm (FA) per hour).}
	\label{tab:KWS_result1}
	\centering
	\footnotesize
	%	\resizebox{\linewidth}{!}{
	\setlength{\tabcolsep}{5.1mm}{
		\begin{tabular}{lccl}
			\toprule
			\textbf{Models}  & \textbf{Params}  & \textbf{FR}  & \textbf{RTF} \\ 
			\midrule
			Baseline \cite{jia20212020}          &   240K          & 2.00  &  0.076 \\
			TDSC-ResNet        &   231K              & 0.72 & 0.043  \\
			\bottomrule
		\end{tabular}
	}
\end{table}
\subsection{Performance of KWS sub-system}

We choose the false rejection rate under one false alarm per hour as the KWS system performance criterion. In a real scenario, the main cost of time comes from the always-on KWS system. We first calculate the process time $T_{kws}$ of KWS system which consists of the time to trigger the system and the time to generate the keyword clips. The real time factor (RTF) is calculated as follows:
\begin{equation}
	\textit{RTF}=T_{kws}/T_{total\_dev}
	\label{eq5}
\end{equation} 
% and here $T_{total\_dev}=20.13$
where $T_{total\_dev}$ is the total duration of the development set audios, and here $T_{total\_dev}=20.13$. The KWS systems are evaluated on an Intel(R) Xeon(R) E5-2643 v4 CPU clocked at 3.4 GHz. As shown in Table~\ref{tab:KWS_result1}, our KWS system has higher accuracy and lower latency compared with the baseline.
\subsection{Performance of SV models}
After KWS system, the keyword segments of the whole utterance in development set compose a new dataset, which will determine the threshold of the following SV system. We choose the mean threshold of Equal Error Rate (EER) and minDCF \cite{greenberg20132012} as the threshold of the SV system.

The statistic $score$ is calculated from the miss rate and the FA rate according to the following equation:
\begin{equation}
	score=Miss+alpha * \textit{FA}
	\label{eq6}
\end{equation}
%The performance of various SV systems can refer to System S4 to S10
where $ Miss $ represents the proportion of errors in all positive label samples, and $ \textit{FA} $ refers to the rate of errors in all negative label samples. The \textit{alpha} constant is set as 19, which is calculated by the assumption that the probability of the positive samples is 0.05.

The results of different systems are shown in Table~\ref{tab:results}. From System S4 to S10, all SV models apply a cosine scoring except the one in S10, which also uses PLDA scoring with the $ embedding \ far $ in Figure~\ref{fig:multi-task}. As the official baseline selected data on OpenSLR\footnote{http://openslr.org/resources.php} as pre-training data, our x-vector system in S5 performs worse than S4. We observe that all the multi-task learning SV models from S6 to S10 achieve better performance than S5. It proves the point that the frame-level phonetic information obtained by multi-task benefits SV system. We choose the MT-CTC-XV-char SV system and use PLDA to estimate scores, then the cosine scores and PLDA scores of S10 are fused. Compared with SV-baseline, the single SV system in S10 gets better performance without other pre-training data. To make SV models more robust, MUSAN\cite{snyder2015musan} and RIRs\cite{ko2017a} noises are added to the training data as a method of data augmentation (S11). Finally, with the score fusion on SV systems in S11 and SV-baseline (S4), we get a competitive result (S12) compared with the top 2 teams (S2,S3), and the score gap on the evaluation set may stem from the additional information of other pre-training data.

%\footnotetext[3]{https://www.pvtc2020.org/leaderboard.html}
\subsection{Effect of multi-task learning}
Meanwhile, different types of multi-task learning are also investigated. Compared with MT-CE-XV (S6), MT-CTC-XV (S7,S8) trains with the label sequence directly and automatically utilizes phonetic information. We find that S8 outperforms S7, and the difference between them is the CTC modeling unit. The Chinese char unit contains more context information compared with phoneme unit, and contributes better performance. However, combining CTC loss and CE loss in multi-CE-CTC (S9) does not lead to further improvement, suggesting that the two phonetic branches may have learned duplicate frame-level phonetic information.

\section{Conclusions}
This paper presents a two-pass system for personalized voice trigger. We applied TDSC-ResNet to Deep-KWS system and made it more accurate and faster. In addition, various multi-task learning framework are explored to jointly learn frame-level phonetic information in SV system. What's more, a MT-CTC-XV system, which is simpler than MT-CE-XV system, is proposed to utilize the phonetic information for SV task. The system can be easily applied to transfer learning and get further improvement. After score fusion, the single MT-CTC-XV sub-system gets a better performance compared with official SV system. At last, the fusion system achieves a score of 0.098 on evaluation set.

\section{Acknowledgements}

The work is supported by the National Natural Science Foundation of China (Grant No.61876160 and No.62001405).

\bibliographystyle{IEEEtran}

\bibliography{mybib}

% \begin{thebibliography}{9}
% \bibitem[1]{Davis80-COP}
%   S.\ B.\ Davis and P.\ Mermelstein,
%   ``Comparison of parametric representation for monosyllabic word recognition in continuously spoken sentences,''
%   \textit{IEEE Transactions on Acoustics, Speech and Signal Processing}, vol.~28, no.~4, pp.~357--366, 1980.
% \bibitem[2]{Rabiner89-ATO}
%   L.\ R.\ Rabiner,
%   ``A tutorial on hidden Markov models and selected applications in speech recognition,''
%   \textit{Proceedings of the IEEE}, vol.~77, no.~2, pp.~257-286, 1989.
% \bibitem[3]{Hastie09-TEO}
%   T.\ Hastie, R.\ Tibshirani, and J.\ Friedman,
%   \textit{The Elements of Statistical Learning -- Data Mining, Inference, and Prediction}.
%   New York: Springer, 2009.
% \bibitem[4]{YourName17-XXX}
%   F.\ Lastname1, F.\ Lastname2, and F.\ Lastname3,
%   ``Title of your INTERSPEECH 2021 publication,''
%   in \textit{Interspeech 2021 -- 20\textsuperscript{th} Annual Conference of the International Speech Communication Association, September 15-19, Graz, Austria, Proceedings, Proceedings}, 2020, pp.~100--104.
% \end{thebibliography}

\end{document}